\documentclass[a4paper,aps,preprintnumbers,twocolumn,prl,showpacs]{revtex4}
\usepackage{amssymb}

\usepackage{bbm}
\usepackage{mathrsfs}
\usepackage{amsmath}
\usepackage[dvips]{graphicx}
\usepackage{epsfig}
\usepackage[dvips]{graphics,graphicx}
\usepackage{color}
\usepackage[dvips]{graphics,graphicx}

\newcommand{\be}{\begin{equation}}
\newcommand{\ee}{\end{equation}}
\newcommand{\bea}{\begin{eqnarray}}
\newcommand{\eea}{\end{eqnarray}}
\newcommand{\bes}{\begin{split}}
\newcommand{\ees}{\end{split}}

\begin{document}

\title{ Superfluidity and collective oscillations of trapped Bose-Einstein
condensates in a periodical potential.}
\author{C. Trallero-Giner}
\affiliation{Faculty of Physics, Havana University, 10400 Havana, Cuba}
\affiliation{Departamento de Fisica, Universidade Federal de S\~{a}o Carlos, 13.565-905,
S \~{a}o Carlos, S\~{a}o Paulo, Brazil}
\author{V. L\'{o}pez-Richard}
\affiliation{Departamento de Fisica, Universidade Federal de S\~{a}o Carlos, 13.565-905,
S \~{a}o Carlos, S\~{a}o Paulo, Brazil}
\author{Y. N\'{u}\~{n}ez-Fern\'{a}ndez}
\affiliation{Faculty of Physics, Havana University, 10400 Havana, Cuba}
\author{Maurice Oliva}
\affiliation{Faculty of Physics, Havana University, 10400 Havana, Cuba}
\author{G. E . Marques}
\affiliation{Departamento de Fisica, Universidade Federal de S\~{a}o Carlos, 13.565-905,
S \~{a}o Carlos, S\~{a}o Paulo, Brazil}
\author{Ming-Chiang Chung}
\affiliation{Physics Division, National Center for Theoretical Science, Hsinchu, 30013,
Taiwan}
\date{\today }

\begin{abstract}
Based on a unified theoretical treatment of the 1D Bogoliubov-de Genes
equations, the superfluidity phenomenon of the Bose-Einstein condensates
(BEC) loaded into trapped optical lattice is studied. Within the
perturbation regime, an all-analytical framework is presented enabling a
straightforward phenomenological mapping of the collective excitation and
oscillation character of a trapped BEC where the available experimental
configurations also fit.
\end{abstract}

\maketitle

\section{Introduction}

Harmonically trapped Bose-Einstein condensates (BECs) offer a great chance
to understand the macroscopic quantum phenomena such as phase coherence~\cite
{Anderson, Hagley, Tosi} and matter wave diffraction~\cite{Ovchinnikov}.
Condensates loaded in a periodic potential forming an optical lattice (OL)~
\cite{Jaksch, Greiner} may show a rich dynamic picture of oscillations as
Bloch oscillations~\cite{Raizen, Choi}, Belieav and Landau damping~\cite
{Katz,Ferlaino}, Landau-Zeeman tunneling~\cite{Raizen}, and the appearance
of the superfluid oscillation of condensates~\cite{Kagan,Inguscio}. Although
schemes of controlling the dynamics of BECs have been profusely described~
\cite{Choi,Chaos3,Chaos4,Berry}, the underlying physics of some of these
studies appears hidden under numerical analysis. The description in terms of
excited states or Goldstone modes not only highlights the main cause of this
behavior, but also enables the characterization of the BEC dynamics in
universal terms. Thus, we select a platform of the BEC loaded simultaneously
into a harmonic traps and an optical lattice to characterize the phenomenon
of superfluidity as well as the dynamical properties and tackled the problem
analytically. The purpose of this letter is to derive a perturbative
treatment which allows explicit closed solutions for the phonon dispersion
relation and to reveal the effect of BEC configurations on the superfluidity
phenomenon. The method has undergone the test of comparison with
experimental evidences with success.

\begin{figure}[tbp]
\includegraphics[scale=1.4]{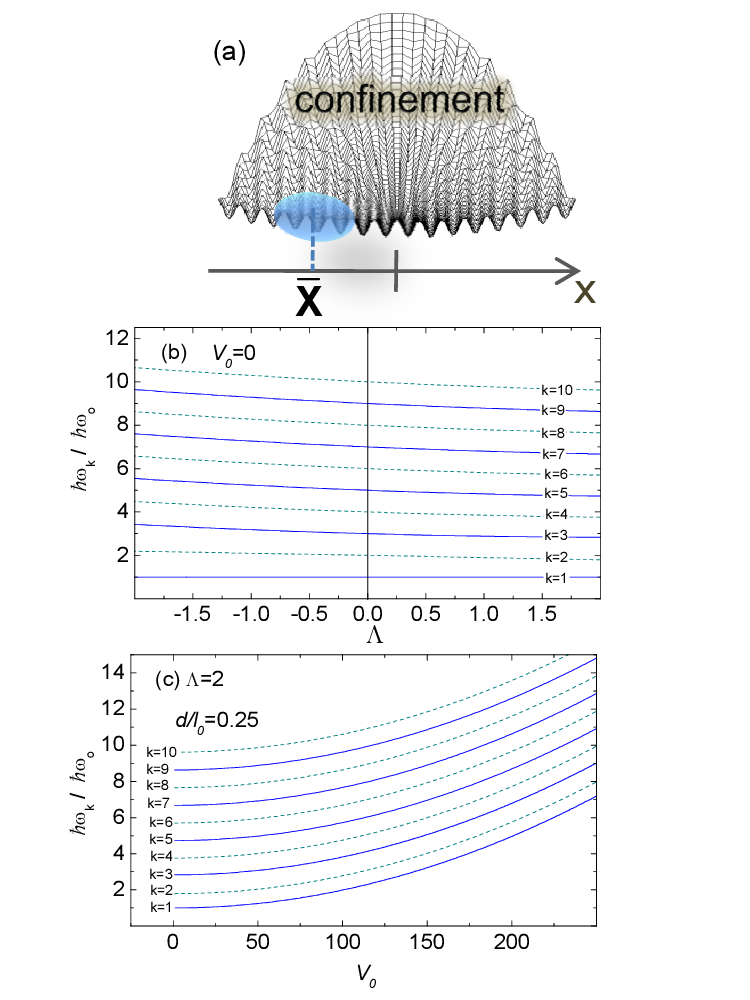}
\caption{(Color online). (a) The oscillating BEC loaded into a 1D optical
lattice within a parabolic trap (grid). The elliptical spot symbolizes the
condensate with a center of mass, $\overline{X}(t)$, oscillating around the
trap bottom. (b) Collective excitation energies $\hbar \protect\omega _{k}$
of the first 10 modes calculated as function of dimensionless parameters $
\Lambda =\protect\lambda _{1D}/(l_{o}\hbar \protect\omega _{0})$ for $V_{o}=0
$ , and (c) as function of $V_{o}=V_{L}/\hbar \protect\omega _{0}$ for $
\Lambda =2$ and $d/l_{o}=0.25$. The solid (dashed) lines represent odd
(even) modes.}
\label{fig1}
\end{figure}

Systems such as the cigar-shaped trap schematically represented in the upper
panel of Fig.~\ref{fig1}, can be considered quasi-one-dimensional (1D)
confinement. Within the framework of mean field theory, the physical
characteristics of a BEC loaded in such trapping profile are ruled by the
time dependent nonlinear Gross-Pitaevskii equation (GPE)~\cite{Gross}.

\begin{equation}
i\hbar \partial _{t}\left\vert \Psi \right\rangle =\left[ -\frac{\hbar ^{2}}{
2m}\frac{\partial ^{2}}{\partial x^{2}}+V_{ext}+\lambda _{1D}\left\vert
\left\vert \Psi \right\rangle \right\vert ^{2}\right] \left\vert \Psi
\right\rangle ,  \label{GPT}
\end{equation}%
where $\lambda _{1D}$ is the self-interaction parameter, $m$ is the alkaline
atom mass, $V_{ext}(x)=\frac{1}{2}m\omega _{0}^{2}x^{2}-V_{L}\cos ^{2}\left(
\dfrac{2\pi }{d}x\right) $ represents the harmonic trap potential plus the
periodic potential caused by the counter-propagating lasers with $V_{L}$ the
laser intensity, $d$ its laser wavelength, and $\omega _{0}$ the frequency
of the harmonic trap. For Eq.~(\ref{GPT}), it is possible to prove
rigorously , the existence of ground states for any $\lambda _{1D}$ and $
V_{L}$. Moreover, the set of ground states is orbitally stable and the
ground states have a Gaussian-like exponential asymptotic behavior for any $
\lambda _{1D}$, regardless of $V_{L}$ value ~\cite{Cazenave}. An important
consequence of this mathematical fact is the stability of those physical
magnitudes, such as no explosion and no damping as function of time, which
are described by operators defined in the Hilbert space of the 1D GPE (\ref
{GPT}). This result is particularly related to the superfluidity properties,
among others physical phenomena, of the harmonically confined condensates
loaded in optical lattices.
\

The collective excitations, or so-called Goldstone modes of the BEC, can be
obtained by applying a small deviation from the stationary solutions $
\left\vert \Psi _{0}\right\rangle $ of Eq.~(\ref{GPT}),
\begin{eqnarray}
\left\vert \Psi (t)\right\rangle &=&\exp (-i\mu t/\hslash )\left[ \left\vert
\Psi _{0}\right\rangle +\left\vert u\right\rangle \exp (-i\omega t)\right.
\notag \\
&&+\left. \left\vert u\right\rangle \exp (-i\omega t)+\left\vert v^{\ast
}\right\rangle \exp (i\omega t)\right] ,  \label{Psi}
\end{eqnarray}
which corresponds to linearizing the time-dependent nonlinear Sch\"{o}dinger
equation in terms of amplitudes $\left\vert u\right\rangle $ and $\left\vert
v^{\ast }\right\rangle $, $\mu $ being the chemical potential and $\omega $
the mode or phonon frequencies~\cite{Ruprecht}. Inserting $\left\vert \Psi
(t)\right\rangle $ into Eq.~(\ref{GPT}), we obtain the Bogoliubov-de Gennes
equations
\begin{equation}
\left[
\begin{array}{cc}
\mathcal{L} & \lambda _{1D}\left\vert \Psi _{0}\right\rangle ^{2} \\
&  \\
-\lambda _{1D}\left\vert \Psi _{0}\right\rangle ^{2} & -\mathcal{L}
\end{array}
\right] \left[
\begin{array}{c}
\left\vert u\right\rangle \\
\\
\left\vert v\right\rangle%
\end{array}
\right] =\hbar \omega _{k}\left[
\begin{array}{c}
\left\vert u\right\rangle \\
\\
\left\vert v\right\rangle
\end{array}
\right] ,  \label{B2}
\end{equation}
where $\mathcal{L}=\widehat{p}^{2}/2m+V_{ext}-\mu +2\lambda
_{1D}\left\langle \Psi _{0}\right. \left\vert \Psi _{0}\right\rangle $.
Three coupled nonlinear equations for $\left\vert \Psi _{0}\right\rangle $, $
\left\vert u\right\rangle $ and $\left\vert v\right\rangle $ must be solved
simultaneously, which incorporate the harmonic trap and the stationary
optical potential. In general, this is a very onerous task and it is not
always possible to extract transparent solutions giving reliable information
on the BEC dynamics. If the harmonic trap potential is switched off from Eqs.
~(\ref{GPT}) and (\ref{B2}) it is possible analytically extract reliable
information of the 1D condensate in a periodic potential as the Bloch
oscillation and stability of the solution~ \cite{Wu,Baroni}. This
corresponds to the homogeneous case where the phonon wavevector $\mathbf{q}
=q_{x}\mathbf{e}_{x}$ is a good quantum number and the excited frequency $
\omega _{\mathbf{q}}=\omega (q)$ is a continuous function of $\mathbf{q}$.
For $\omega _{0}\neq 0$ the wavevector $\mathbf{q}$ is no longer a good
quantum number (inhomogeneous case) and the system (\ref{GPT} - \ref{B2})
provides a set of discrete excited states $\omega _{k}$ labeled by $
k=1,2,....$ By assuming a weakly-interacting Bose gas and not too strong
laser intensities, the self-induced nonlinear interaction and the OL
potential can be considered as perturbations with respect to the harmonic
trap potential. Accordingly, compact solutions for $\mu $ and the spatial
shape of the order parameter $\left\vert \Psi _{0}\right\rangle ,$ as
determined by relevant parameters of the condensate are obtained~\cite
{trallero3}.

\section{Normal modes}

Considering the nonlinear term $\lambda _{1D}\left\langle \Psi _{0}\right.
\left\vert \Psi _{0}\right\rangle $ and the periodical potential $V_{L}\cos
^{2}\left( \dfrac{2\pi }{d}x\right) $ as small terms compared to the
confined harmonic trap potential strength, the solutions of the system\ (\ref
{B2}) can be cast in terms of the complete set of harmonic oscillator wave
functions $\{\left\vert \psi _{n}\right\rangle \}$~\cite{nota}: $\left\vert
u\right\rangle =\sum_{n=0}^{\infty }A_{n}\left\vert \psi _{n}\right\rangle $
and\ $\left\vert v\right\rangle =\sum_{n=0}^{\infty }B_{n}\left\vert \psi
_{n}\right\rangle $ and the coefficients $A_{n}$ and $B_{n}$ are expanded in
form of series $A_{n}=A_{n}^{(1)}+A_{n}^{(2)}+...;$ $
B_{n}=B_{n}^{(1)}+B_{n}^{(2)}+...,$ where the quantities $A_{n}^{(i)}$ and $
B_{n}^{(i)}$ $\sim $ $\Lambda ^{i}$\textbf{\ , }$V_{o}^{i}$. Using Eqs.~(\ref
{B2}) we get

\begin{eqnarray}
&&\sum\limits_{n,i}A_{n}^{(i)}\left( B_{n}^{(i)}\right) \left[ \left( k+
\frac{1-V_{o}}{2}-\frac{\mu }{\hbar \omega _{0}}-\left( +\right) \frac{
\omega _{k}}{\omega _{0}}\right) \delta _{k,n}\right.  \notag \\
&&\left. -\frac{1}{2}V_{o}\left\langle \psi _{k}\right\vert \cos 2\alpha
x\left\vert \psi _{n}\right\rangle +2\Lambda \left\langle \psi _{k}\right.
\left\vert \Psi _{0}\right\rangle \left\langle \Psi _{0}\right\vert \left.
\psi _{n}\right\rangle \right]  \notag \\
&=&-\Lambda \sum\limits_{n,i}B_{n}^{(i)}\left( A_{n}^{(i)}\right)
\left\langle \psi _{k}\right. \left\vert \Psi _{0}\right\rangle \left\langle
\Psi _{0}\right\vert \left. \psi _{n}\right\rangle ,  \label{s1}
\end{eqnarray}
with $\alpha =2\pi l_{o}/d,$ $l_{o}=\sqrt{\hbar /m\omega _{0}},$ $\Lambda
=\lambda _{1D}/(l_{o}\hbar \omega _{0}),$ $V_{o}=V_{L}/\hbar \omega _{0}$.
Taking advantage of the procedure developed in Ref.~\onlinecite{trallero3}
and solving simultaneously the system\ (\ref{s1}), it is possible to show
that the independent phonon frequencies $\omega _{k}$ are given by

\begin{eqnarray}
\frac{\omega _{k}}{\omega _{0}} &=&k+\frac{\Lambda }{\sqrt{2\pi }}\left[ -1+
\frac{2\Gamma (k+1/2)}{\sqrt{\pi }k!}\right] -  \notag \\
&&\frac{V_{o}}{2}\exp \left( -\alpha ^{2}\right) \left[ L_{k}(2\alpha
^{2})-1 \right] -  \notag \\
&&\frac{\Lambda V_{o}}{\sqrt{2\pi }}\exp \left( -\alpha ^{2}\right) \left[
Ei(\frac{\alpha ^{2}}{2})-\mathcal{C}-\ln \frac{\alpha ^{2}}{2}+\frac{\delta
_{k}(\alpha )}{\sqrt{\pi }}\right] +  \notag \\
&&\frac{V_{o}^{2}}{4}\exp \left( -2\alpha ^{2}\right) \left[ Chi(2\alpha
^{2})-\mathcal{C-}\ln 2\alpha ^{2}+\rho _{k}(\alpha )\right]  \notag \\
&&+\Lambda ^{2}\left[ \frac{\gamma _{k}}{2\pi ^{2}}+0.033106\right] \text{
,\ \ \ \ \ }k=1,2,...,  \label{phonon}
\end{eqnarray}
where $L_{k}(z),$ $\Gamma (z)$, $Ei(z),$ $Chi(z)$, and $\mathcal{C}$ are the
Laguerre polynomials, the gamma function, the exponential integral, the
cosine hyperbolic integral, the Euler's constant, respectively. Finally, $
\gamma _{k}$ are numbers, $\delta _{k}$ and $\rho _{k}$ being explicit
functions the dimensionless parameter $\alpha $~\cite{nota1,nota2}.

In Fig.~\ref{fig1} the analytical solutions for the frequencies $\omega _{k}$
are graphically represented for the first 10 modes. Panel~(b) displays $
\omega _{k}$ for attractive ($\Lambda <0$) and repulsive ($\Lambda >0$)
cases at $V_{0}=0.$ Also in panel (c) the influence of the periodical
potential on the phonon modes being checked for fixed values $d/l_{0}=0.25$
and $\Lambda =2.$ In the first case, the collective oscillations show an
almost flat dispersion as a function $\Lambda .$ Note that the mode for $k=1$
has the frequency value $\omega _{0}$ of the harmonic trap~\cite{pitaevskii}
. In panel (c) is also seen a blue-shift renormalization of $\omega _{k}$
can be noted due to the presence of the OL.

The normalized eigenvectors $\left\vert \Phi _{k}\right\rangle ^{\dagger
}=\left( \left\vert u_{k}^{\ast }\right\rangle ,\left\vert
v_{k}\right\rangle \right) $ can be cast as

\begin{equation}
\left\vert \Phi _{k}\right\rangle =\left(
\begin{array}{c}
\left\vert \psi _{k}\right\rangle +\sum\limits_{m\neq k}\dfrac{\left(
4\Lambda f_{k,m}-V_{o}g_{k,m}\right) }{2(k-m)}\left\vert \psi
_{m}\right\rangle \\
-\sum\limits_{m=0}^{\infty }\dfrac{\Lambda f_{k,m}}{k+m}\left\vert \psi
_{m}\right\rangle%
\end{array}
\right) ,  \label{Fik}
\end{equation}
with $f_{k,m}=\left( -1\right) ^{(k-m)/2}\Gamma \left( \frac{k+m+1}{2}
\right) /(\pi \sqrt{2m!k!}),$ $g_{k,m}=\left( -1\right) ^{(k-m)/2}p!\left(
2\alpha \right) ^{\left\vert k-m\right\vert }L_{p}^{\left\vert
k-m\right\vert }(2\alpha ^{2})2^{p}/(\sqrt{2^{k+m}m!k!})\times \exp \left(
-\alpha ^{2}\right) $, $L_{p}^{t}(z)$ being the Laguerre polynomials, $
p=(k+m-\left\vert k-m\right\vert )/2$ and $m+k=$ even number. Notice that
the quasiparticle amplitudes $\left\vert u_{k}\right\rangle $ and $
\left\vert v_{k}^{\ast }\right\rangle $ present parity inversion symmetry
property, i.e. if the index $k$ is an even or an odd number we are in
presence of two independent subspaces where the wave functions $\left\vert
\Phi _{k}\right\rangle $ become symmetric (even mode) or antisymmetric (odd
mode) with respect the transformation $x\rightarrow -x.$.

According to Eqs.~(\ref{Psi})\ and (\ref{Fik}), for a given time $t,$ the
probability density $\left\vert \left\langle x\right\vert \left. \Psi
_{k}(t)\right\rangle \right\vert ^{2}$ of the excited states along the $x$
-axis shows oscillations with well defined maxima which are quenched
according to the exponential behavior $\exp \left( -x^{2}/l_{o}^{2}\right) .$
The position of the maxima\ and minima of the axial density $\left\vert
\left\langle x\right\vert \left. \Psi _{k}(t)\right\rangle \right\vert ^{2}$
are linked to the minima or maxima of the combined potential function $
V/\hslash \omega _{0}=$ $0.5(x/l_{o})^{2}-V_{o}\cos ^{2}\left( 2\pi
x/l_{o}\right) $. Finally, another important limit is reached when the
optical lattice is turned off. In this case the standard solutions (in a
perturbative sense) of Bogoliubov-de Gennes equations for the inhomogeneous
case are directly obtained from Eqs.~(\ref{phonon}) and (\ref{Fik}) taking $
V_{o}=0$ .

\section{Superfluidity.}

\begin{figure}[tbp]
\includegraphics[scale=0.40]{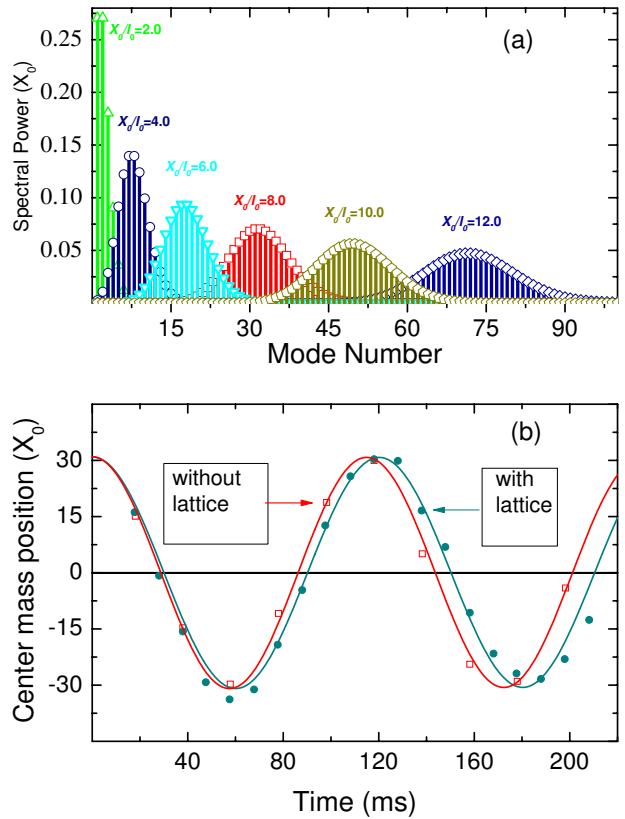}
\caption{(Color online) (a) Universal spectral power contributions to the
BEC oscillations within the perturbation regime. The inset shows the mode
number, $k_{m}$, where the maximum value of the spectral power is attained.
(b) Center-of-mass position. Experimental data from Ref.~
\onlinecite{Inguscio} are represented by open squares and full circles,
while by solid lines show the calculations after Eqs.~(\protect\ref{phonon})
and ( \protect\ref{XO}).}
\label{fig2}
\end{figure}

To characterize the dynamics of a BEC loaded in an OL we evaluated the
expectation value of the center-of-mass position $\overline{X(t)}$ =$
\left\langle \Psi (t)\right\vert x\left\vert \Psi (t)\right\rangle ,$ with $
\left\vert \Psi (t)\right\rangle =\sum\limits_{k=0}^{\infty }C_{k}\left\vert
\phi _{k}(t)\right\rangle .$ Here, the set of eigensolutions is chosen as $
\left\vert \phi _{0}(t)\right\rangle =\exp (-i\mu t/\hslash )\left\vert \Psi
_{0}\right\rangle ,$ $\left\vert \phi _{k}(t)\right\rangle =\exp (-i\mu
t/\hslash )\left[ \exp (-i\omega _{k}t)\left\vert u_{k}\right\rangle +\exp
(i\omega _{k}t)\left\vert v_{k}^{\ast }\right\rangle \right] $ for $k\neq 0,$
and the coefficients \{$C_{k}\}$ are obtained under certain initial
condition. In our case, we consider that, at $t=0$, the OL is absent and
condensate\ is located out of equilibrium at certain distance $X_{0}$ from
the origin, i.e. the order parameter $\left\vert \Psi _{0}\right\rangle $ is
centered at $x=X_{0}$ with an expectation value $\overline{X(0)}=X_{0}$. At,
$t>0$ the system may or may not be loaded into the OL periodical potential.

A straightforward calculation, by keeping terms up to first-order in $
\Lambda $ and $V_{0}$,\ yields that the dynamics of the center-of-mass is
ruled by the equation $\overline{X(t)}=X_{0}(t)+X_{\Lambda ,V_{0}}(t),$ where

\begin{equation}
X_{0}(t)=X_{0}\exp \sum\limits_{k=0}^{\infty }F_{k}\cos \left( \omega
_{k+1}-\omega _{k}\right) t,  \label{XO}
\end{equation}
$F_{k}=\exp (-X_{0}^{2}/2l_{o}^{2})\cdot X_{0}^{2k}/[\left(
2l_{o}^{2}\right) ^{k}k!]$ is the spectral power and $X_{\Lambda ,V_{0}}(t)$
is a linear function of $\Lambda $ and$V_{0},$ with negligible contribution
to $\overline{X(t)}$ for typical experimental setups$.$ In Eq.~\ref{XO} the
condition $\omega _{k=0}=0$ is used.

The spectral power contribution to these oscillations, in terms of the mode
number, $k$, exclusively dependent on the relative initial position, $
X_{0}/l_{0}$, as displayed in Fig.~\ref{fig2} (a). For a given initial
displacement, the mode $k=k_{m}$ with maximal contribution to the $X_{0}(t)$
is given by the equation $\ln \left( X_{0}^{2}/2l_{o}^{2}\right) -H_{k_{m}}-
\mathcal{C}=0,$ with $H_{n}=\sum_{k=1}^{n}1/k$. Thus, one can see the larger
the relative initial displacement, $X_{0}/l_{0}$, the higher the main
contributing modes, as shown in the inset of Fig.~\ref{fig2} (a), and wider
their diffusion. This has energetic implications since the mode frequencies,
$\omega _{k}$, obtained from Eq.~(\ref{phonon}) are tunable with the
nonlinear interaction strength and the OL parameters. Without OL the
vibrational level spacing $\Delta \omega _{k}=\omega _{k+1}-\omega
_{k}\approx \omega _{0}$ which make the expectation value $
X_{0}(t)=X_{0}\cos \omega _{0}t$~\cite{pitaevskii}.

Figure~\ref{fig2} (b) shows the superfluid oscillation for a BEC of $^{87}$
Rb extracted from Ref.~\onlinecite{Inguscio} measured in a static magnetic
trap with and without a 1D periodic potential, as simulated by our approach.
Using Eqs.~(\ref{phonon}) and (\ref{XO}) we are able to reproduce the
reported experimental center-of-mass position as function of time without
using any fitting parameter. The oscillations observed in Fig.~\ref{fig2}
(b) correspond to vibrational level spacing $\Delta \omega _{k}$ of the two
independent subspaces with even and odd modes that make $X(t)$ oscillate
with a frequency near the harmonic value $\omega _{0}.$ The small frequency
shift of the condensate loaded into OL,\ and observed in Fig.~\ref{fig2}
(b), is directly linked to the renormalization of the atomic mass of the
system moving in a periodical potential.

In summary, we presented a unified analytical description for the collective
excitations (phonon frequencies $\omega _{k},$ Eq.~(\ref{phonon}), the
excited states wavefunctions $\left\vert \Phi _{k}\right\rangle ,$ Eq.~(\ref
{Fik})), the 1D superfluidity oscillation and the dynamics (center-of-mass
position $X_{0}(t)$ Eq.~(\ref{XO})) of BEC systems loaded in an optical
lattice.

This work was partially supported by Alexander von Humboldt Foundation. C.
T.-G., V. L-R. and G. E . M. acknowledge the financial support of Brazilian
agencies FAPESP, CAPES, and CNPq. M. C. Chung acknowledges NSC in Taiwan. C.
T-G. is grateful to J. M. Rost for many insightful discussions and the
hospitality enjoyed during his stay at the Max-Planck-Institut f\"{u}r
Physik Komplexer Systeme.


\begin{thebibliography}{99}
\bibitem{Anderson} Anderson B. P. and Kasevich M. A. , \textit{Science},
\textbf{282}, (1998) 1686.

\bibitem{Hagley} Hagley E. W. \textit{et al.}, \textit{Science}, \textbf{283}
, (1999) 1706.

\bibitem{Tosi} Chiofalo M. L. and Tosi M. P. , \textit{Phys. Lett. A}
\textbf{268}, (2000) 406.

\bibitem{Ovchinnikov} Ovchinnikov Y. B. \textit{et al.}, \textit{Phys. Rev.
Lett.} \textbf{83}, (1999) 284.

\bibitem{Jaksch} Jaksch D. \textit{et al.}, \textit{Phys. Rev. Lett.}
\textbf{81}, (1998) 3108.

\bibitem{Greiner} Greiner M. \textit{et al.,} \textit{Nature (London)}
\textbf{415}, (2002) 39.

\bibitem{Raizen} Raizen M. \textit{et al.}, \textit{Phys. Today} \textbf{50}
, (1997) 30.

\bibitem{Choi} Choi D.-I. and Niu Q., \textit{Phys. Rev. Lett. }\textbf{82},
(1999) 2022.

\bibitem{Katz} Katz N. \textit{et al.}, \textit{Phys. Rev. Lett.} \textbf{89}
, (2002) 220401.

\bibitem{Ferlaino} Ferlaino F. \textit{et al.}, \textit{Phys. Rev. A}
\textbf{66}, (2002) 011604.

\bibitem{Inguscio} Burger S., \textit{et al., Phys. Rev. Lett. }\textbf{86},
(2001) 4447.

\bibitem{Kagan} Kagan Yu. and Maksimov L. A., \textit{Phys. Rev. Lett.}
\textbf{85}, (2000) 3075.

\bibitem{Chaos3} Brezinova I., \textit{et al.}, \textit{Phys. Rev. A}
\textbf{83}, (2011) 043611.

\bibitem{Chaos4} Wanga Z., \textit{et al.}, \textit{J. Exp. Theor. Phys.}
\textbf{112}, (2011) 355.

\bibitem{Berry} Berry N. H. and Kutz J. N., \textit{Phys. Rev. E}, \textbf{%
75 } (2007) 036214.

\bibitem{Gross} Gross E. P., \textit{Nuovo Cimento} \textbf{20}, (1961) 454
; Pitaevskii L. P. , \textit{Zh. Eksp. Teor. Fiz.} \textbf{40}, (1961) 646 [
\textit{Sov. Phys. JETP} \textbf{13}, (1961) 451].

\bibitem{Cazenave} Cazenave T. and Lions P. L., \textit{Commun. Math. Phys}.
\textbf{85}, (1982) 549; R. Cipolatti, \textit{et al.,.} arXiv:1107.2704v1,
(2011).

\bibitem{Ruprecht} Ruprecht P. A., \textit{et al.}, \textit{Phys. Rev. A},
\textbf{54}, (1996) 4178.

\bibitem{Wu} Wu B. and Niu Q., \textit{Phys. Rev. A}, \textbf{64}, (2001)
061603.

\bibitem{Baroni} Barontini G, and Modugno M. \textit{Phys. Rev. A,} \textbf{%
\ 76}, 041601 (2007)

\bibitem{trallero3} Trallero-Giner C., \textit{et al.},\textit{Phys. Rev. A}
, (2009) 063621.

\bibitem{nota} For a detailed description of the stationary solution $%
\left\vert \Psi _{0}\right\rangle ,$ the chemical potencial and the validity
of the perturbative method see Ref.~\onlinecite{trallero3}.

\bibitem{nota1} In typical experiments the values of the dimensionless
parameter $\alpha =2\pi l_{o}/d$ ranges between 10 and 50 (see Morsch O. and
Oberthaler M., \textit{Rev. Mod. Phys.} \textbf{78}, (2006) 179). This
allows a simplification of the reported analytical expresions for the
chemical potential $\mu $ and also for the excited frequencies $\omega _{k}$.

\bibitem{nota2}
\begin{widetext}
\begin{eqnarray*}
\gamma _{k} &=&\frac{2}{\sqrt{\pi }}\sum\limits_{m\neq 0}\frac{\left(
-1\right) ^{m+1}\Gamma ^{2}(m+.5)\Gamma (k-m+.5)}{mm!k!2^{2m}}-\frac{\Gamma
^{2}(k+0.5)}{2k(k!)^{2}}-\sum\limits_{m\neq k}\left( \frac{1}{m+k}+\frac{4}{m-k}\right) \frac{
\Gamma ^{2}(\frac{m+1+k}{2})}{m!k!},
\end{eqnarray*}
\end{widetext}
\begin{widetext}
\begin{eqnarray*}
\delta _{k}(\alpha ) &=&\frac{2}{\pi }\sum\limits_{m\neq 0}\frac{\left(
-2\right) ^{m-1}\alpha ^{2m}\Gamma ^{2}(m+.5)\Gamma (k-m+.5)}{m\left(
2m\right) !k!}+2(-1)^{k}\sum\limits_{m\neq k}\frac{\left( p\right) !\left( 2\alpha
\right) ^{\left\vert m-k\right\vert }\Gamma (\frac{m+1+k}{2})}{\left(
2\right) ^{\frac{\left\vert m-k\right\vert }{2}}m!k!(k-m)}L_{p}^{\left\vert
k-m\right\vert }(2\alpha ^{2}),
\end{eqnarray*}
\end{widetext}
and%
\begin{equation*}
\rho _{k}(\alpha )=\sum\limits_{m\neq k}\frac{\left( p!\right) ^{2}\left(
2\alpha ^{2}\right) ^{\left\vert m-k\right\vert }}{m!k!(k-m)}\left[
L_{p}^{\left\vert k-m\right\vert }(2\alpha ^{2})\right] ^{2},
\end{equation*}
where $\sum $ is restricted to $m+k=$ even number,\ $p=(k+m-\left\vert
k-m\right\vert )/2,$ and $L_{p}^{t}(z)$ are the Generalized Laguerre
polinomials.

\bibitem{pitaevskii} Pitaevskii L. and Stringari S. ,\textit{Bose-Einstein
Condensation}, (Clarendon Press, Oxford) 2003.
\end{thebibliography}
\end{document}